\documentclass[prl,fleqn,showpacs,twocolumn]{revtex4-1}
\usepackage{graphicx}
\usepackage{amsmath}
\usepackage{subfigure}
\begin{document}

\title{Selfish atom selects quantum resonances at fractional atomic frequencies}
\author{Gennady~A. Koganov*}
\email{gkoganov@gmail.com}
\affiliation{Physics Department, Ben Gurion University of the Negev, P.O.Box 653, Beer Sheva, 84105, Israel}

\author{Reuben Shuker}
\email{reuben.shuker@gmail.com}
\affiliation{Physics Department, Ben Gurion University of the Negev, P.O.Box 653, Beer Sheva, 84105, Israel}

\begin{abstract}
We show that the atom as a "quantum entity", driven by an external field in the form of pulse sequence at repetition rate equal to the internal quantum frequency divided by an integer n, responds resonantly. It seeks and finds its characteristic frequencies in any possible combination of its frequencies. This is an indication of self expression by the atom at many sub-frequencies of its own transition frequencies. It is a non-intuitive phenomenon since the external repetition rate has no quantum character, yet the atom responds to it if the rate is equal to 1/n its eigen-frequency. We believe that our results will have implications in other quantum related processes, such as resonant enhancement of chemical reactions and biological processes.
\end{abstract}

\maketitle

When a quantum system is driven by a periodic external field the interaction can have a unique resonant character.  A few examples: a dipole interaction of a two-level system with resonant electromagnetic field \cite{Alen-Eberly}, the phenomenon of nuclear magnetic resonance \cite{NMR} where nuclei in a magnetic field are in resonance with external electromagnetic field, electron paramagnetic resonance \cite{EPR} where the frequency of the driving field equals to frequency separation between the two energy levels of unpaired free electrons, etc. Resonance phenomena are widely and successfully used in precise measurements, various kinds of spectroscopy, generating coherent electromagnetic field and so on. 

In this Letter we propose a new concept of unique pulsed quantum resonances, based on an idea of driving a quantum system by an external field in the form of a repetitive pulse sequence. In this situation one would intuitively expect a resonant behavior when the pulse repetition rate is close to the frequency separation between two particular quantum states of the system. Less intuitive and, probably striking finding is the appearance of additional resonances at fractional frequencies equal $1/n$ of the main resonant frequency.  The theoretical study described here is motivated by a recent experiment \cite{experiment}. The $1/n$ theoretical result is in good qualitative agreement with the experiment. We have also found that other resonances are possible at pulse repetition rate frequencies equal to $1/n$ of combination frequencies involving the internal system frequencies and  Rabi frequencies.  We show that quantum interference related phenomena of gain without population inversion and absorption with inversion are observed at these unique resonances. We expect that such a resonant response may be exploited in studying various phenomena by applying pulse repetitive magnetic and electric fields on the Zeeman and the Stark manifolds, respectively. This work has ramifications, beyond the atomic level to enhance quantum chemical and biological processes, using pulse repetitive interaction.

\begin{figure}[htbp]
\centering
\includegraphics[scale=0.6]{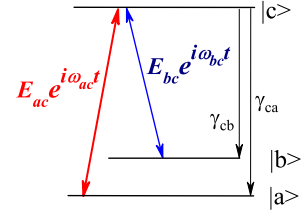}
\caption{(Color online) $\Lambda$- like three-level scheme. Frequency separation between the lower states $\left|a\right\rangle$ and$\rightarrow\left|b\right\rangle$ is small enough, so that applied fields can interact with both $\left|b\right\rangle\rightarrow\left|c\right\rangle$ and $\left|a\right\rangle\rightarrow\left|c\right\rangle$ transitions.}\label{Scheme}
\end{figure}

We consider, as an example, a tree-level system shown in Fig. \ref{Scheme} driven by two laser fields, $E_{ac}$ and $E_{bc}$, which are tuned in resonance (however, not restricted to exact resonance) with the frequencies $\omega_{ac}$ and $\omega_{bc}$ of the atomic transitions $\left|c\right\rangle\rightarrow\left|a\right\rangle$ and $\left|c\right\rangle\rightarrow\left|b\right\rangle$, respectively.

The frequency separation $\omega_{ab}$ between the two lower states $\left|a\right\rangle$ and $\left|b\right\rangle$ is considered to be small compared to those between the upper state $\left|c\right\rangle$ and both lower states $\left|a\right\rangle$ and $\left|b\right\rangle$, so that both $\left|c\right\rangle\rightarrow\left|a\right\rangle$ and $\left|c\right\rangle\rightarrow\left|b\right\rangle$ transitions can be driven by the same pumping laser. One can, for example, think of these two levels as Zeeman sub-levels due to a weak magnetic field. The system Hamiltonian in the interaction picture and the rotating wave approximation has the following form:

{\raggedright
\begin{eqnarray}
H=-(\hat{d}_{ac}\cdot\stackrel{\rightarrow}{E}_{ac}f_{1}(t)+\hat{d}_{bc}\cdot\stackrel{\rightarrow}{E}_{bc}f_{2}(t)+ \nonumber \\ 
\hat{d}_{ac}\cdot\stackrel{\rightarrow}{E}_{bc}f_{2}(t)e^{-i\omega_{ab}t}+\hat{d}_{bc}\cdot\stackrel{\rightarrow}{E}_{ac}f_{1}(t)e^{i\omega_{ab}t})+H.c. ,
\label{Hamiltonian}
\end{eqnarray}
}
%\begin{equation}
%H=-\hbar [\Omega_{ac} e^{i\Delta_{ac}t}(\sigma_{ac}+e^{i\omega_{ab}t}\sigma_{bc})f(t)+\Omega_{bc} e^{i\Delta_{bc}t}(\sigma_{bc}+e^{-i\omega_{ab}t}\sigma_{ac})]+H.c. ,
%\label{Hamiltonian2}
%\end{equation}

\noindent where $\hat{d}_{ac}=d_{ac}\sigma_{ac}$ and $\hat{d}_{bc}=d_{bc}\sigma_{bc}$ are dipole operators of the corresponding transitions, $\sigma_{ac}=\left|c\rangle\langle a\right|$ and $\sigma_{bc}=\left|c\rangle\langle b\right|$ are atomic raising/lowering operators. The first two terms in (\ref{Hamiltonian}) represent resonant interaction between the driving fields $E_{ac}$ and $E_{bc}$ and the corresponding transitions. The last two terms describe cross-interaction between the driving field $E_{ac}$ ($E_{bc}$) and the transition  $\left|b\right\rangle\rightarrow\left|c\right\rangle$ ( $\left|a\right\rangle\rightarrow\left|c\right\rangle$), in which case the fields are automatically detuned by the atomic frequency $\pm\omega_{ab}$ from the corresponding transitions. Functions  $f_{1}(t)$ and $f_{2}(t)$ which determine the time dependencies of the driving field amplitudes $E_{ac}$ and $E_{bc}$, respectively, are chosen either as 1, in the case of cw field, or as $\sum g(t-n\tau)$ in the case of a sequence of short pulses, or any their combination. The function g here is taken Gaussian, but any appropriate pulse function can be used as well.  The Hamiltonian (\ref{Hamiltonian}) contains four Rabi frequencies which can be defined as following: $\Omega_{ac}=\stackrel{\rightarrow}{d}_{ac}\cdot\stackrel{\rightarrow}{E}_{ac}/\hbar$, $\Omega_{bc}=\stackrel{\rightarrow}{d}_{bc}\cdot\stackrel{\rightarrow}{E}_{bc}/\hbar$, $\Omega_{acb}=\stackrel{\rightarrow}{d}_{bc}\cdot\stackrel{\rightarrow}{E}_{ac}/\hbar$, and $\Omega_{bca}=\stackrel{\rightarrow}{d}_{ac}\cdot\stackrel{\rightarrow}{E}_{bc}/\hbar$. These frequencies, in combination with the low frequency $\omega_{ab}$ of the $\left|a\right\rangle\rightarrow\left|b\right\rangle$ transition determine the possible resonances in the system. 

To find these quantum resonances we have solved numerically the time-depended semiclassical master equation with Hamiltonian (\ref{Hamiltonian}) (see, for example, \cite{Doron-Dressed}).  Semiclassical equations of motion for the atomic density matrix elements read:

\begin{widetext}
\begin{flalign}
& \dot{\rho}_{aa}  =   \gamma_{ab}\rho_{bb}+\gamma_{ac}\rho_{cc}- i\Omega_{ac}f_{1}(t)(\rho_{ac}-\rho_{ca}) -  if_{2}(t)(\Omega_{bca}e^{-i\omega_{ab}t}\rho_{ac} -\Omega_{acb}e^{i\omega_{ab}t}\rho_{ca}) \label{raa}\\
&  \dot{\rho}_{bb}  =   -\gamma_{ab}\rho_{bb}+\gamma_{bc}\rho_{cc}- i\Omega_{bc}f_{2}(t)(\rho_{bc}-\rho_{cb}) +  i\Omega_{acb}f_{1}(t)(e^{i\omega_{ab}t}\rho_{bc} -e^{-i\omega_{ab}t}\rho_{cb}) \label{rbb}\\
& \dot{\rho}_{bc}=-\frac{\gamma_{ac}+\gamma_{bc}+\gamma_{ab}}{2}\rho_{bc}-i[\Omega_{ac}f_{1}(t)+e^{i\omega_{Lar}t}\Omega_{bca}f_{2}(t)]\rho_{ba}+ i[e^{-i\omega_{Lar}t}\Omega_{acb}f_{1}(t)+\Omega_{bc}f_{2}(t)](\rho_{cc}-\rho_{bb}) \label{rbc} \\
& \dot{\rho}_{ac}=-\frac{\gamma_{ac}+\gamma_{bc}}{2}\rho_{ac}-i[\Omega_{bc}f_{2}(t)+e^{-i\omega_{Lar}t}\Omega_{acb}f_{1}(t)]\rho_{ab}+ i[e^{i\omega_{Lar}t}\Omega_{bca}f_{2}(t)+\Omega_{ac}f_{1}(t)](\rho_{cc}-\rho_{aa}) \label{rac}\\
& \dot{\rho}_{ab}=-\frac{\gamma_{ab}}{2}\rho_{ab}-i[\Omega_{bc}f_{2}(t)+e^{i\omega_{Lar}t}\Omega_{acb}f_{1}(t)]\rho_{ac}+i([\Omega_{ac}f_{1}(t)+e^{i\omega_{Lar}t}\Omega_{bca}f_{2}(t)]\rho_{cb} \label{rab}
\end{flalign}
\end{widetext}

\begin{figure}[h]
\centering
\includegraphics[scale=0.3]{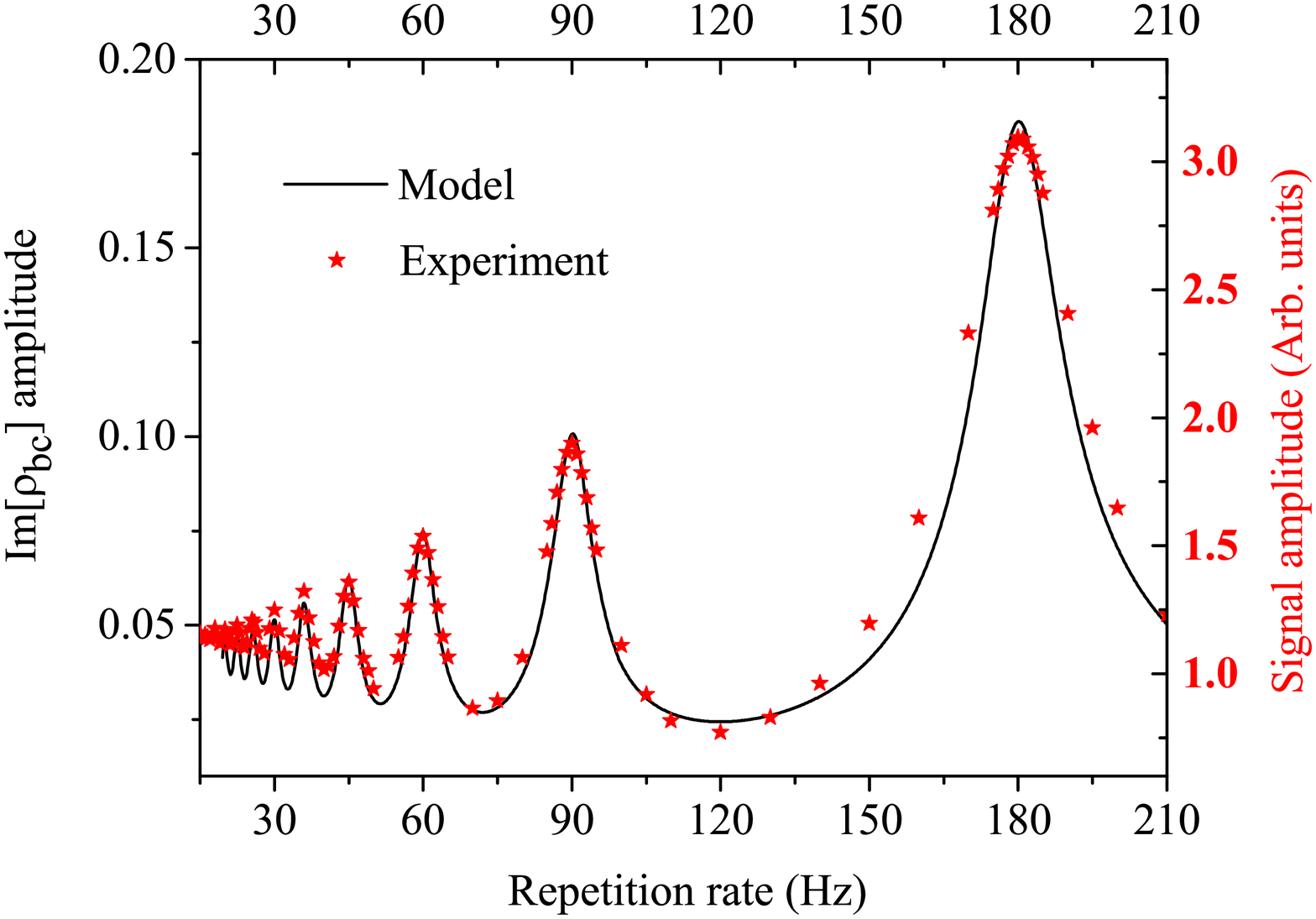}
\caption{(Color online) Oscillation amplitude of the absorption on the probe transition as function of the pulse repetition rate $\omega_{Rep}$ (solid line). Apart from expected resonance at $\omega_{Rep}=\omega_{ab}$ there are additional resonances at $\omega_{Rep}=\omega_{ab}/n$, where n is an integer number. Experimental data from \cite{experiment} (stars).}\label{Exper}
\end{figure}

We calculated the atomic coherences and populations at various repetition rates $\omega_{Rep}=2\pi/\tau$ of the driving pulse sequence. As a first example we take cw $E_{ac}$ and pulsed $E_{bc}$. Figure \ref{Exper} (solid line) shows dependence of oscillation amplitude of $Im[\rho_{bc}]$ responsible for absorption on the probe transition $\left|c\right\rangle\rightarrow\left|b\right\rangle$, on the repetition rate $\omega_{Rep}$. One can see that in addition to intuitively expected main resonance at $\omega_{Rep}=\omega_{ab}$, there appear other resonances at fractional frequencies $\omega_{Rep}=\omega_{ab}/n$ with $n$ being an integer number. This brings one to conclude that among all possible pulse repetition frequencies $\omega_{Rep}$ the system selects those which are in resonance not only with its eigen-frequency $\omega_{ab}$, but also with its subharmonics $\omega_{ab}/n$. \\

Indeed, similar fractional frequency resonances have been observed recently in experiment with atomic magnetometer \cite{experiment}, where the pulse repetition rate of the pumping laser was in resonance with the frequency separation $\omega_{Lar}$ between Zeeman sub-levels (Larmor frequency) and its sub-harmonics $\omega_{Lar}/n$. Figure \ref{Exper} (red stars) exhibits the experimental signal amplitude as a function of the pump repetition rate. As can be seen, there is a good qualitative agreement between the experimental data and theoretical calculations from the model shown in Fig. \ref{Scheme}. 

Appearance of the described above resonances, being surprising at first glance, can be understood looking at a simple classical analogy: equation of motion of an harmonic oscillator with eigen-frequency $\omega_{0}$, driven by external force in the form of short pulse sequence, reads

\begin{equation}\label{oscillator}
\stackrel{..}{x}+b\stackrel{.}{x}+\omega^2_{0}x=\sum_{n}{\delta(t-n\tau)},
\end{equation}

\noindent where b is the decay rate due to friction or any other dissipation process. The solution of Eq.(\ref{oscillator}) has resonances at $\omega=\omega_{0}/n$, where $\omega=2\pi/\tau$ is pulse repetition frequency. A simple example from every day life is provided by a kids swing: when you give a push periodically, at proper time moments, the swing reaches some stationary amplitude. What would happen if you skip every second cycle of the swing, so that the repetition frequency of your force is half the swing eigen-frequency? the swing again will be in resonance, but with lower amplitude. Then if you skip every third, fourth etc. cycles, the system will go into resonance corresponding to $\omega=\omega_{0}/n$ with less and less amplitude. If we plot the oscillation amplitude obtained from the solution of Eq. (\ref{oscillator}) as a function of the force repetition frequency $\omega$, it will look very similar to the solid line in Fig. \ref{Exper}. \\

\begin{figure}[h]
\centering
\subfigure[]{\label{populations}
\includegraphics[scale=0.3]{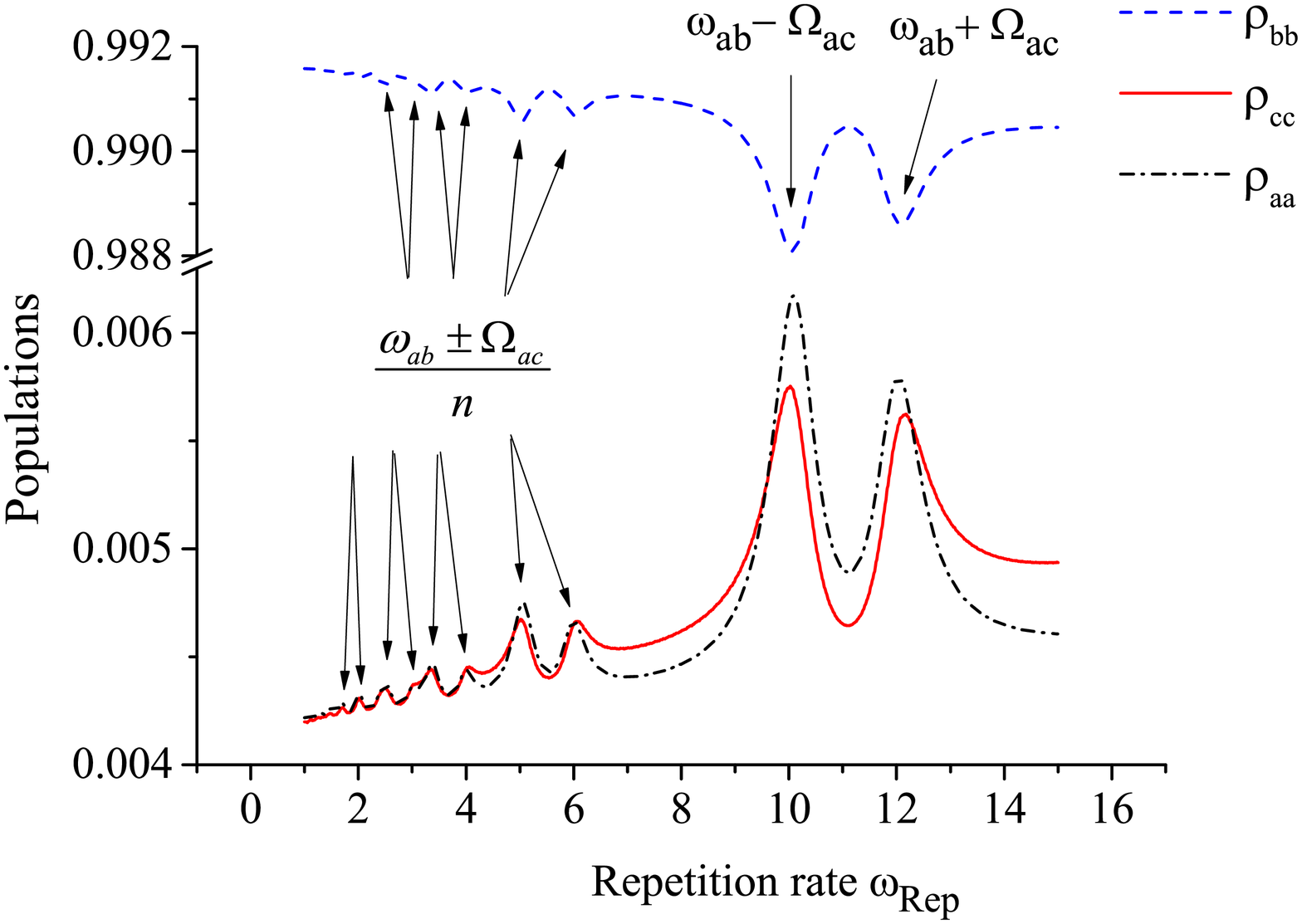}}
\subfigure[]{\label{coherences}
\includegraphics[scale=0.3]{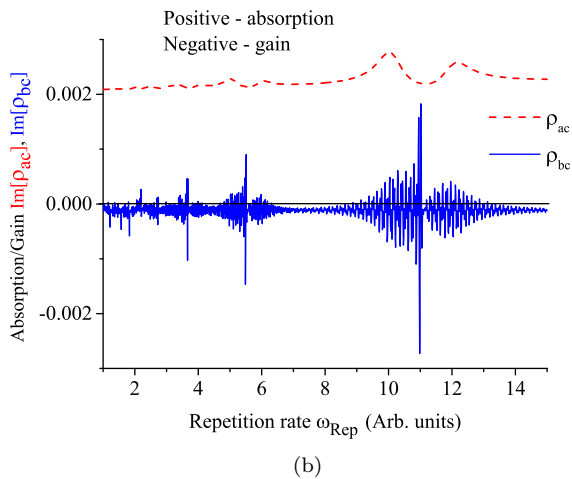}}
\caption{(Color online) Combination resonances at repetition rates $\omega_{Rep}=\left|\Omega_{ac}\pm \omega_{ab}\right|/n$ and quantum-interference-related phenomena. (a) Populations of atomic states: CPT at all repetition rates. (b) Absorption/gain on the pump (dashed red line) and probe (solid blue line) transitions: at appropriate values of the pulse repetition rate we have GWI on the probe transition  $\left|b\right\rangle\rightarrow \left|c\right\rangle$ at combination resonances $\omega_{Rep}=\left|\Omega_{ac}\pm \omega_{ab}\right|/n$, ADI on the pump transition  $\left|a\right\rangle\rightarrow \left|c\right\rangle$ out of combination resonances. $\Omega_{ac}=1, \omega_{ab}=11$.}\label{combination}
\end{figure}

\begin{figure}[h]
\centering
\includegraphics[scale=0.3]{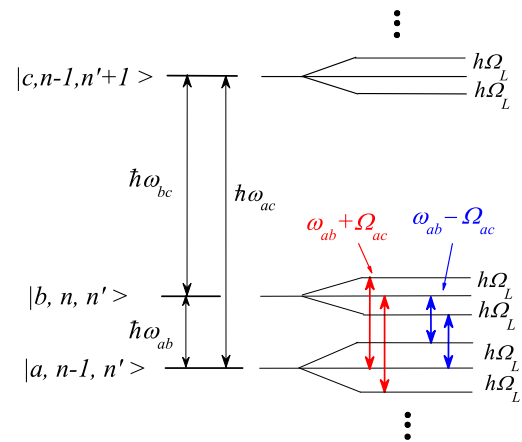}
\caption{(Color online) Dressed states of the atom
+ laser photon system as calculated along the line presented in Ref. \cite{Doron-Dressed}. Left (red online) and right (blue online) couples  of arrows show the main resonances at sum and difference frequencies $\omega_{ab}+\Omega_{ac}$ and $\omega_{ab}-\Omega_{ac}$, respectively.}\label{dressed}
\end{figure}

In the next example we demonstrate what we call ``combination resonances''. To this end we allow the field $E_{ac}$ interacting with $\left|a\right\rangle\rightarrow\left|c\right\rangle$ transition to have both cw and pulse components, while another field $E_{bc}$ is absent. Figure \ref{combination} shows populations of the atomic states (Fig. \ref{populations}) and absorption/gain on the pump and probe transitions (Fig. \ref{coherences}) as functions of the repetition rate $\omega_{Rep}$. As it can be observed, the main resonances occur at the combination frequencies $\left|\omega_{ab}\pm\Omega_{ac}\right|$. This can be readily understood if we look at the dressed states \cite{Cohen-Tanudji} picture shown in Fig. \ref{dressed}, in which a partial dressed states manifold is shown for the case of  $\omega_{ab} > \Omega_{ac}$ \cite{Doron-Dressed}. Again, apart from the main resonance there appear additional fractional resonances, this time at combination frequencies $\left|\omega_{ab}\pm\Omega_{ac}\right|/n$.

\begin{figure}[h]
\centering
\includegraphics[scale=0.3]{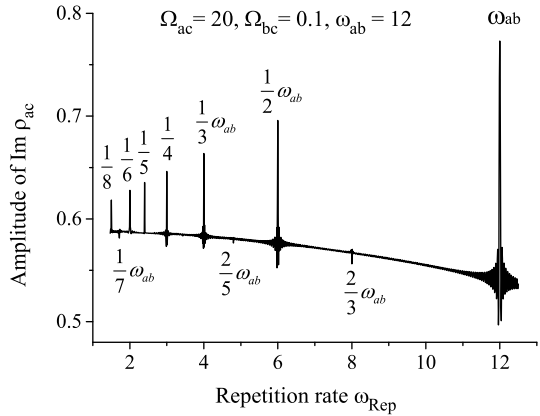}
\caption{Strong pulsed pump and weak cw probe fields. Apart form $\omega_{Rep}=\omega_{ab}/n$ resonances, additional resonances appear at rational frequencies $m\omega_{ab}/n$, in this case at $1/2, 2/5$ and $2/3$.}\label{R02-narrow}
\end{figure}

Figure \ref{combination} also demonstrates quantum-interference-related phenomena, accompanying combination resonances, namely, Coherent Population Trapping (CPT) \cite{CPT-Review,Arimondo1996257}, Lasing/Gain Without Inversion (LWI/GWI) \cite{EIT-Review},  and Absorption Despite Inversion (ADI) \cite{Scully-PhysRevLett.62.2813,Doron-AWI-2001}. Indeed, one can observe from Fig. \ref{populations} that the intermediate state $\left|b\right\rangle$ is mostly populated at all values of the repetition rate $\omega_{Rep}$, whereas two other states $\left|a\right\rangle$ and $\left|c\right\rangle$ are nearly empty. This is clear manifestation of the phenomenon of coherent population trapping.  On Fig. \ref{coherences} we observe that the absorption on the pump transition $\left|a\right\rangle\rightarrow\left|c\right\rangle$ is always positive, however, in some regions of the pulse repetition rate $\omega_{Rep}$, far from combination resonances $(\omega_{ab}\pm\Omega_{ac})$, this transition is inverted, as can be seen in Fig. \ref{populations}. Hence in those regions we have ADI. The probe transition $\left|b\right\rangle\rightarrow\left|c\right\rangle$ behaves in a different manner, exhibiting GWI except for narrow ranges of the repetition rate $\omega_{Rep}$ near the combination resonances, where exchange between absorption and gain occurs. All the above phenomena are manifestation of quantum interference.

Finally, increasing the pulsed pump Rabi frequency $\Omega_{ac}$ from 1 to 20 and adding weak cw probe $\Omega_{bc}=0.1$, we obtain narrow quantum resonances, as shown in  Fig. \ref{R02-narrow}. In this case all resonances have sharp profiles. Also there are additional fractional resonances, having small amplitudes,  at rational frequencies.

\textit{In summary}, we have shown an unintuitive phenomenon where by atom exhibits a  resonant behavior if an interaction with it has the form of repetitive pulses at repetition rate equals to one of its quantum internal eigen-frequencies divided by an integer. It is remarkable that the atom and indeed practically any quantum system, responds resonantly to an external degree of freedom (such as the frequency of the repetition rate of the excitation), that apparently has nothing to do with the internal quantum degree of freedom. In this respect, the atom is indeed selfish: it exploits these external non-quantal degrees of freedom to respond resonantly. We envisage that such a resonant response may also be achieved applying magnetic field on the Zeeman and electrical field on the Stark manifolds, by including corresponding terms into the system Hamiltonian.
This work has ramifications, beyond the atomic level, on quantum chemical and biological processes. In these, one can exploit repetitive pulsed light or other type of interaction at 
repetition rate frequencies not only equal, but also at $1/n$ of the internal quantum transition frequencies. We hope that these results encourage further theoretical and especially experimental studies in various quantum systems.
%\acknowledgements{}

\bibliographystyle{unsrt}
\bibliography{Koganov}

\begin{thebibliography}{10}

\bibitem{Alen-Eberly}
{L. Allen and J.H. Eberly}.
\newblock {\em Optical Resonance and Two Level Atoms}.
\newblock Wiley, 1975.

\bibitem{NMR}
E.~R. {Andrew}.
\newblock {\em {Nuclear Magnetic Resonance}}.
\newblock Cambridge University Press, 2009.

\bibitem{EPR}
{J.~A. Weil} and J.~R. Bolton.
\newblock {\em {Electron Paramagnetic Resonance: Elementary Theory and
  Practical Applications}}.
\newblock Wiley-Interscience, 2007.

\bibitem{experiment}
R.~Shuker et.al.
\newblock {\em Personal communication}, 2013.

\bibitem{Doron-Dressed}
D~Braunstein and R~Shuker.
\newblock Dressed states analysis of lasing without inversion in a three-level
  ladder system: the steady-state regime.
\newblock {\em Journal of Physics B: Atomic, Molecular and Optical Physics},
  42(12):125401, 2009.

\bibitem{Cohen-Tanudji}
{Claude Cohen-Tanudji, Jacques Dupont-Roc and Gilbert Grynberg}.
\newblock {\em Atom-Photon interactions: Basic Processes and Applications}.
\newblock John Wiley, Chapter VI, 1992.

\bibitem{CPT-Review}
B~D Agap'ev, M~B Gornyĭ, B~G Matisov, and Yu~V Rozhdestvenskiĭ.
\newblock Coherent population trapping in quantum systems.
\newblock {\em Physics-Uspekhi}, 36(9):763, 1993.

\bibitem{Arimondo1996257}
E.~Arimondo.
\newblock V coherent population trapping in laser spectroscopy.
\newblock volume~35 of {\em Progress in Optics}, pages 257 -- 354. Elsevier,
  1996.

\bibitem{EIT-Review}
Michael Fleischhauer, Atac Imamoglu, and Jonathan~P. Marangos.
\newblock Electromagnetically induced transparency: Optics in coherent media.
\newblock {\em Rev. Mod. Phys.}, 77:633--673, Jul 2005.

\bibitem{Scully-PhysRevLett.62.2813}
Marlan~O. Scully, Shi-Yao Zhu, and Athanasios Gavrielides.
\newblock Degenerate quantum-beat laser: Lasing without inversion and inversion
  without lasing.
\newblock {\em Phys. Rev. Lett.}, 62:2813--2816, Jun 1989.

\bibitem{Doron-AWI-2001}
D.~Braunstein and R.~Shuker.
\newblock Absorption with inversion and amplification without inversion in a
  coherently prepared v system: A dressed-state approach.
\newblock {\em Phys. Rev. A}, 64:053812, Oct 2001.

\end{thebibliography}

\end{document}